\documentclass[letterpaper,conference,10pt]{IEEEtran}

\usepackage{etex}
\usepackage{cite}

\usepackage{graphicx}
\usepackage{multirow}

\usepackage[cmex10]{amsmath}
\usepackage{amssymb}
\usepackage{nicefrac}

\usepackage{fixltx2e}

\usepackage{booktabs}
\usepackage{dcolumn}
\usepackage{tabularx}

\usepackage{ifpdf}
\ifpdf
\usepackage[draft,pdfborder={0 0 0}]{hyperref}
\fi

\usepackage{bm}
\usepackage[varg]{txfonts}
\let\mathbb=\varmathbb
\DeclareSymbolFont{letters}{OML}{ztmcm}{m}{it}

\usepackage{tikz,pgf}
\usepackage{pgfplots}
\usetikzlibrary{shapes,positioning,arrows,decorations.markings,fit,calc,patterns}

\usepackage{algorithm}
\usepackage{algorithmicx}
\usepackage{algpseudocode}
\algrenewcommand{\algorithmiccomment}[1]{\textbf{//}#1}

\usepackage[font=normalsize]{subfig}
\usepackage{color}
\newcommand{\fixme}[2]{\ifx&#2&{\color{red}#1}\else{\color{red}FIXME\{}#1{\color{red}\}}\footnote{{\color{red}#2}}\PackageWarning{Fixme}{#1: #2}\fi}

\usepackage[final]{changes}

\newcommand{\mvec}[1]{\bm{#1}}

\newcolumntype{C}[1]{>{\centering\arraybackslash}m{#1}}
\newcolumntype{R}[1]{>{\raggedleft\arraybackslash}m{#1}}

\title{Blind Detection of Polar Codes}

\author{\IEEEauthorblockN{Pascal Giard, %
Alexios Balatsoukas-Stimming, and %
Andreas Burg}
  \IEEEauthorblockA{Telecommunications Circuits Laboratory,\\\'Ecole polytechnique f\'ed\'erale de Lausanne (EPFL), Lausanne, Switzerland.\\Email: \{pascal.giard,alexios.balatsoukas,andreas.burg\}@epfl.ch}
  }

\usepackage{glossaries}

\usepackage{flushend}
\newcommand{\metricfigheight}{5.70cm}
\newcommand{\ecperffigheight}{6.85cm}

\begin{document}
\newacronym{ue}{UE}{user-equipment}
\newacronym{dci}{DCI}{downlink control information}
\newacronym{pdcch}{PDCCH}{physical downlink control channel}
\newacronym[plural=CCEs,firstplural=control channel elements (CCEs)]{cce}{CCE}{control channel element}
\newacronym{rnti}{RNTI}{radio network temporary identifier}
\newacronym[plural=UESSS,firstplural=UE-specific search spaces (UESSS)]{uesss}{UESSS}{UE-specific search space}
\newacronym[plural=CSS,firstplural=common search spaces (CSS)]{css}{CSS}{common search space}
\newacronym{tti}{TTI}{transmission time interval}
\newacronym[plural=LLRs,firstplural=log-likelihood ratios (LLRs)]{llr}{LLR}{log-likelihood-ratio}
\newacronym{sc}{SC}{successive-cancellation}
\newacronym{spc}{SPC}{single-parity-check}
\newacronym{ber}{BER}{bit-error rate}
\newacronym{fer}{FER}{frame-error rate}
\newacronym[plural=CDFs,firstplural=cumulative distribution functions (CDFs)]{cdf}{CDF}{cumulative distribution function}
\newacronym{lte}{LTE}{Long-Term Evolution}
\newacronym{far}{FAR}{false-alarm rate}
\newacronym[plural=CRCs,firstplural=cyclic-rendundency checks (CRCs)]{crc}{CRC}{cyclic-rendundency check}
\newacronym{ldpc}{LDPC}{low-density parity-check}
\newacronym{bch}{BCH}{Bose-Chaudhuri-Hocquenghem}
\newacronym{roc}{ROC}{receiver operating characteristic}
\newacronym{bpsk}{BPSK}{binary phase-shift keying}
\newacronym{awgn}{AWGN}{additive white Gaussian-noise}

\maketitle

\begin{abstract}
  Polar codes were recently chosen to protect the control channel information in the next-generation mobile communication standard (5G) defined by the 3GPP. As a result, receivers will have to implement blind detection of polar coded frames in order to keep complexity, latency, and power consumption tractable. As a newly proposed class of block codes, \added{the problem of polar-code} blind detection \added{has received very little attention}. 
  In this work, we propose a low-complexity blind-detection algorithm for polar-encoded frames. We base this algorithm on a novel detection metric with update rules that leverage the a priori knowledge of the frozen-bit locations, exploiting the inherent structures that these locations impose on a polar-encoded block of data. We show that the proposed detection metric allows to clearly distinguish polar-encoded frames from other types of data by considering the cumulative distribution functions of the detection metric, and the receiver operating characteristic. The presented results are tailored to the 5G standardization effort discussions, i.e., we consider a short low-rate polar code concatenated with a CRC.
\end{abstract}

\section{Introduction}
\label{sec:introduction}

In modern mobile communications, \gls{ue} devices receive critical control messages through a control channel. These messages can be placed in various valid locations which form the so-called \textit{search space}.
\added{Within this search space, a \gls{ue} receiver is tasked with the identification of messages addressed to it among the candidate locations.}
Furthermore, these messages are protected by channel codes and \glspl{crc} to notably increase reliability and decrease the \gls{far}. Since the detection search space typically contains over forty candidate locations, it is highly desirable for \gls{ue} receivers to avoid running a complex decoder for a modern error-correcting code on all candidates, i.e., it is preferable to eliminate the majority of the candidates early on to minimize the complexity, latency, and power consumption.

To address this problem in previous mobile communication standards, multiple strategies and algorithms for the blind detection of messages encoded with convolutional codes were proposed, e.g., \cite{Shieh2005,Moosavi2011,Sipila2012,Malladi2012}. Some blind-detection algorithms for other types of codes such as \gls{bch} codes \cite{Zhou2013} or \gls{ldpc} codes \cite{Xia2014} were also devised. However, in the next-generation mobile communication standard (5G) developed by the 3GPP, the control channel will be protected by polar codes~\cite{3GPPRAN187}.

\added{Blind detection of polar codes has been independently researched in \cite{Condo_COML_2017}, where a two-step method that employs the path metric as used in list decoding to elect the best candidates is proposed. That work focuses on fitting within the 5G parameters. Our works are orthogonal, and our proposed detection metric can be used with their method.}

In this paper, we propose a low-complexity blind-detection algorithm for polar-encoded frames based on a novel detection metric. We propose to take advantage of the a priori knowledge of the frozen-bit locations in polar codes of a given rate to update a detection metric based on the resulting constituent-code types. Update rules specific to certain constituent-code types are devised and their rationale is explained. The effectiveness of the detection metric is demonstrated by examining its evolution in various scenarios showing that it can very effectively distinguish polar-encoded frames from random data or noise. This demonstration is done by looking at the \glspl{cdf} of the proposed metric and by drawing the \gls{roc}. It should be noted that although results are provided for a systematic polar code, our proposed approach applies to both systematic and non-systematic polar codes.

\textit{Outline:} The remainder of this paper is organized as follows. Section~\ref{sec:background} provides the necessary background on polar codes. Section~\ref{sec:algo} describes our proposed blind-detection algorithm, and introduces the detection metric at the core of our algorithm along with the various update rules. Complexity considerations as well as limitations are also briefly discussed in Section~\ref{sec:algo}. Section~\ref{sec:results} investigates the effectiveness of our proposed blind-detection method. Finally, Section~\ref{sec:conclusion} concludes this paper.

\section{Polar Codes}
\label{sec:background}

\subsection{Construction}
Polar codes, which are capacity-achieving linear channel codes~\cite{Arikan2009}, are based on the application of a linear transformation to a vector of bits before they are transmitted over a communications channel. Polar codes differ from other commonly used codes in that the highly structured nature of the aforementioned linear transformation enables the use of low-complexity encoding and decoding algorithms. Moreover, the application of this linear transformation has a polarizing effect, in the sense that, in the limit of infinite blocklength $N$, some of the bits can be decoded perfectly while the remaining bits are completely unreliable.

More specifically, the polarizing linear transformation for a polar code of blocklength $N$ can be obtained as
\begin{align}
  \label{eq:encoding}
	\mvec{x} = & \mvec{u}\mvec{F}^{\otimes n},	\qquad \mvec{F} = \begin{bmatrix} 1 & 0 \\ 1 & 1 \end{bmatrix},
\end{align}
where $n \triangleq \log _2 N$ and $\mvec{u}$ is the vector of bits to be transmitted. When using this transformation, it is possible to calculate the reliability of transmission for each $u_i,\, i \in \{0,\hdots,N-1\}$~\cite{Arikan2009,Tal2011a}. In order to construct an $(N,K)$ polar code of rate $R = \nicefrac{K}{N}$, the $K-N$ $u_i$'s corresponding to the least-reliable bit positions are frozen to a value that is known at both the transmitter and the receiver (usually $0$), while the remaining $K$ $u_i$'s are used to transmit information. The bits corresponding to the set of $N-K$ least-reliable positions are called \emph{frozen bits}.

In addition to the matrix form, polar codes can also be represented as a graph. Fig.~\ref{fig:pc16graph} shows such a representation for a $(16, 11)$ polar code, where the frozen-bit and information-bit $u_i$ locations are labeled in light gray and in black, respectively.

Polar codes can either be non-systematic---as calculated with \eqref{eq:encoding} or as illustrated by the graph in Fig.~\ref{fig:pc16graph}---, or systematic as discussed in \cite{Arikan2011}. Systematic polar codes offer a slightly better \gls{ber} than their non-systematic counterparts, while both types share the same \gls{fer}. It was shown in \cite{Sarkis_TCOMM_2016} that systematic encoding could be carried out, using the same generator matrix $\mvec{F}^{\otimes n}$, with low complexity. The method proposed in this work applies to both types of polar codes.

\subsection{Constituent Codes and Representation}

\begin{figure}[t]
  \centering
  \hspace{-10pt}\resizebox{\columnwidth}{!}{\input{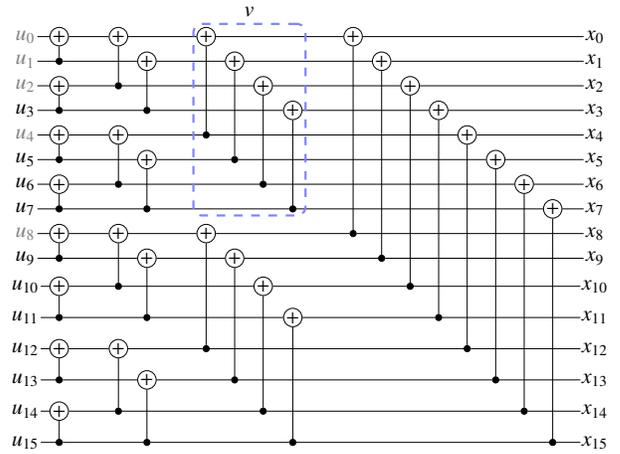}}
  \vspace{-5pt}
  \caption{Graph representation of a $(16, 11)$ polar code.}
  \label{fig:pc16graph}
\end{figure}

Polar codes are built recursivly where at each step, two polar codes of the same length are combined to construct a bigger polar code of twice the length. Consider the combination step circled in blue occuring at $v$ as illustrated in Fig.~\ref{fig:pc16graph}: a polar code of length $N_v=8$ is created by combining 2 polar codes of length $\nicefrac{N_v}{2}=4$, where the first four elements are an element-wise combination---with an exclusive-or (XOR) operation---of the polar codes of length $\nicefrac{N_v}{2}$ and the other four elements are a copy of the elements composing the second polar code. A polar code can be seen as being built out of smaller constituent (polar) codes. Furthermore, by considering the frozen-bit locations, some of these constituent codes specialize as other types of codes~\cite{Alamdar-Yazdi2011,Sarkis_JSAC_2014}, e.g., a polar code where only the most significant location contains an information bit while all the other locations are frozen is a Repetition code.

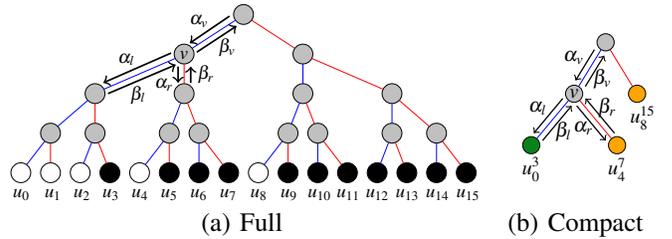
\begin{figure}
  \vspace{-8pt}
  \centering
    \subfloat[Full]{\label{fig:tree-sc-16-11}
      \begin{tikzpicture}[every node/.style={font=\scriptsize}]
        \node[anchor=south west,inner sep=0] (image) at (0,0) {\includegraphics[width=0.7\columnwidth]{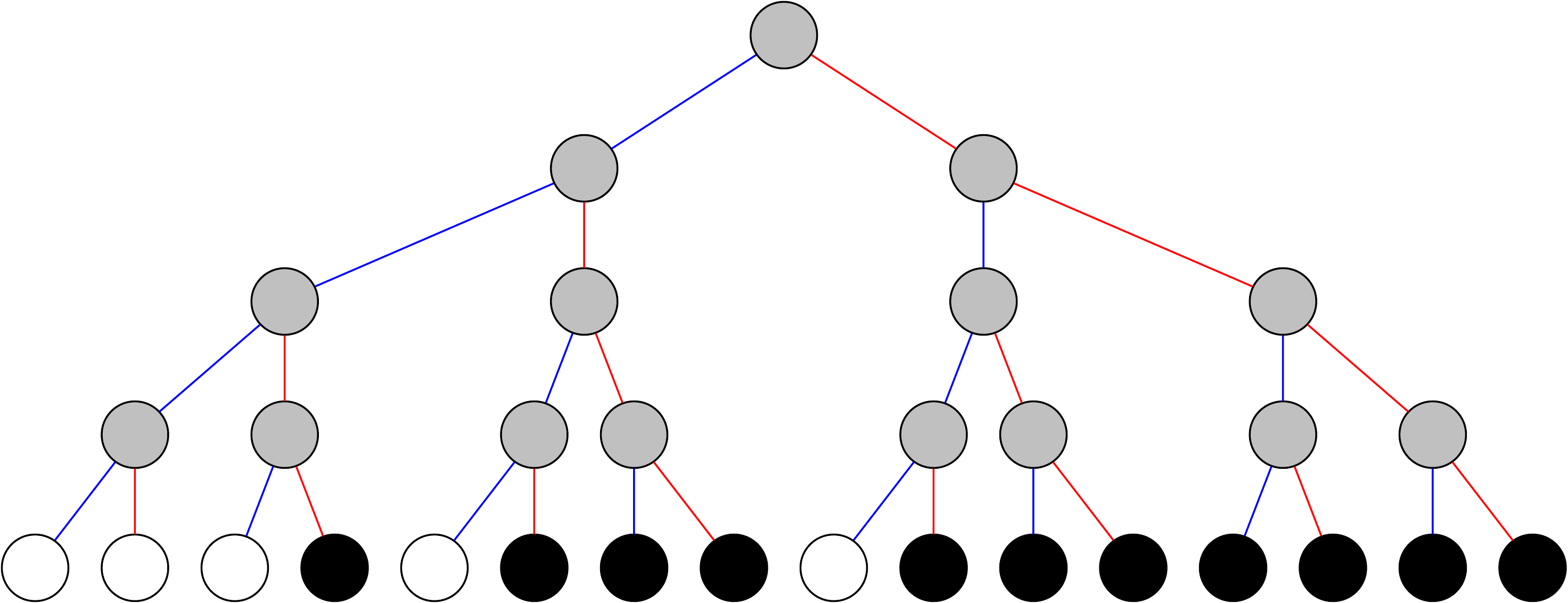}};
        \begin{scope}[x={(image.south east)},y={(image.north west)}]
          \node (v) at (.3725,.715) {$v$};
          \path[->,line width=0.65pt] ($(.47,.93)$) edge ($(.385,.78)$); 
          \path[->,line width=0.65pt] ($(.40,.73)$) edge ($(.485,.875)$); 
          \path[->,line width=0.65pt] ($(.345,.725)$) edge ($(.2,.56)$); 
          \path[->,line width=0.65pt] ($(.21,.5)$) edge ($(.355,.66)$); 
          \path[->,line width=0.65pt] ($(.36,.65)$) edge ($(.36,.56)$); 
          \path[->,line width=0.65pt] ($(.387,.56)$) edge ($(.387,.65)$); 
          \node (av) at (.405,.9) {$\alpha_v$};
          \node (bv) at (.465,.755) {$\beta_v$};
          \node (al) at (.25,.69) {$\alpha_l$};
          \node (bl) at (.275,.495) {$\beta_l$};
          \node (ar) at (.33,.55) {$\alpha_r$};
          \node (br) at (.42,.61) {$\beta_r$};
          \node (u0) at (.025,-.06) {$u_0$};
          \node (u1) at ($(u0)+(0.068,0)$) {$u_1$};
          \node (u2) at ($(u1)+(0.062,0)$) {$u_2$};
          \node (u3) at ($(u2)+(0.062,0)$) {$u_3$};
          \node (u4) at ($(u3)+(0.062,0)$) {$u_4$};
          \node (u5) at ($(u4)+(0.063,0)$) {$u_5$};
          \node (u6) at ($(u5)+(0.063,0)$) {$u_6$};
          \node (u7) at ($(u6)+(0.063,0)$) {$u_7$};
          \node (u8) at ($(u7)+(0.065,0)$) {$u_8$};
          \node (u9) at ($(u8)+(0.065,0)$) {$u_9$};
          \node (u10) at ($(u9)+(0.065,0)$) {$u_{10}$};
          \node (u11) at ($(u10)+(0.062,0)$) {$u_{11}$};
          \node (u12) at ($(u11)+(0.062,0)$) {$u_{12}$};
          \node (u13) at ($(u12)+(0.064,0)$) {$u_{13}$};
          \node (u14) at ($(u13)+(0.065,0)$) {$u_{14}$};
          \node (u15) at ($(u14)+(0.065,0)$) {$u_{15}$};
        \end{scope}
      \end{tikzpicture}}
  \subfloat[Compact]{\label{fig:tree-fastssc-16-11}\quad\resizebox{0.22\columnwidth}{!}{\rotatebox{90}{


\definecolor{deepgreen}{RGB}{8, 130, 25}
\definecolor{mygray}{RGB}{192, 192, 192}
\definecolor{myorange}{RGB}{255, 165, 0}

\begin{tikzpicture}[
        level/.style={level distance = 9mm},
        level 1/.style={sibling distance=11mm, edge from parent/.style={draw,blue,line width=0.5pt}},
        level 2/.style={sibling distance=15mm, edge from parent/.style={draw,blue,line width=0.5pt}},
        ]

\tikzset{
frozen/.style={draw=black,fill=white,minimum size=3mm,circle, inner sep=0},
fullspace/.style={draw=black,fill=black,minimum size=3mm,circle, inner sep = 0},
mixed/.style={draw=black,fill=mygray,minimum size=3mm,circle, inner sep = 0},
rep_mixed/.style={draw=black,fill=deepgreen,minimum size=3mm,circle, inner sep = 0},
spc_mixed/.style={draw=black,fill=myorange,minimum size=3mm,circle, inner sep = 0}
}

\tikzset{
parallel segment/.style={
   segment distance/.store in=\segDistance,
   segment pos/.store in=\segPos,
   segment length/.store in=\segLength,
   to path={
   ($(\tikztostart)!\segPos!(\tikztotarget)!\segLength/2!(\tikztostart)!\segDistance!90:(\tikztotarget)$) -- 
   ($(\tikztostart)!\segPos!(\tikztotarget)!\segLength/2!(\tikztotarget)!\segDistance!-90:(\tikztostart)$)  \tikztonodes
   }, 
   segment pos=.5,
   segment length=5ex,
   segment distance=-1mm,
},
}

\node[mixed] (3_0){} [grow=left]
	child {node[mixed] (2_0){\rotatebox{-90}{$v$}}
  	child {node[rep_mixed] (0_4){}
    }
		child {node[spc_mixed] (1_2){} edge from parent[red]
		}
	}
	child {node[spc_mixed] (2_1){} edge from parent[red]
	}
;

\draw[->,line width=0.65pt] (3_0) to[parallel segment,segment length=4ex] node[above right=-1.5mm] {\rotatebox{-90}{$\alpha_v$}} (2_0);
\draw[->,line width=0.65pt] (2_0) to[parallel segment] node[above right=-1.5mm] {\rotatebox{-90}{$\alpha_l$}} (0_4);
\draw[->,line width=0.65pt] (0_4) to[parallel segment] node[below left=-2.25mm] {\rotatebox{-90}{$\beta_l$}} (2_0);
\draw[->,line width=0.65pt] (2_0) to[parallel segment] node[above left=-2.25mm] {\rotatebox{-90}{$\alpha_r$}} (1_2);
\draw[->,line width=0.65pt] (1_2) to[parallel segment] node[below right=-2.25mm] {\rotatebox{-90}{$\beta_r$}} (2_0);
\draw[->,line width=0.65pt] (2_0) to[parallel segment,segment length=4ex] node[below left=-2.25mm] {\rotatebox{-90}{$\beta_v$}} (3_0);

\node at ($(0_4)-(.42,0)$) {\rotatebox{-90}{$u_0^3$}};
\node at ($(1_2)-(.42,0)$) {\rotatebox{-90}{$u_4^7$}};
\node at ($(2_1)-(.42,.1)$) {\rotatebox{-90}{$u_8^{15}$}};

\end{tikzpicture}
}}}
  \vspace{-3pt}
  \caption{Decoder-tree representation of a (16, 11) polar code.}
  \label{fig:tree-16-11}
\end{figure}

To make their representation more compact, it was proposed to represent polar codes as binary trees (or decoder trees)~\cite{Alamdar-Yazdi2011,Sarkis_JSAC_2014}. Fig.~\ref{fig:tree-16-11} shows two decoder-tree representations of the polar code reprensented as a graph in Fig.~\ref{fig:pc16graph}. Fig.~\ref{fig:tree-sc-16-11} is a direct translation of the graph into a decoder tree, where each leaf node is either a frozen-bit location (white) or an information-bit location (black). Fig.~\ref{fig:tree-fastssc-16-11} is an even more compact representation where the leaf nodes are constituent codes: $u_0^3$ \added{(green) is a Repetition code, and $u_4^7$ and $u_8^{15}$ (orange) are both \gls{spc} codes}.

\subsection{Decoding}

To decode polar codes, algorithms traverse either one of the decoder trees illustrated in Fig.~\ref{fig:tree-16-11}. Algorithms taking advantage of the a priori knowledge of the frozen-bit locations traverse a decoder tree like the one of Fig.~\ref{fig:tree-fastssc-16-11} while the others traverse the one of Fig.~\ref{fig:tree-sc-16-11}. Specifically, it was shown in \cite{Sarkis_JSAC_2014} that a polar code can be efficiently decoded, in terms of speed, by decomposing it in smaller constituent codes of different types and by using dedicated decoding algorithms on them. That algorithm, called fast-SSC, was shown to match the error-correction performance of the original \gls{sc} algorithm while significantly reducing latency and increasing throughput. What remains the same, however, is that in all cases, decoding takes place by traversing the decoder tree depth first starting with the root node and moving along the left edge (blue) first.

In \cite{Tal2015}, it was proposed to build a constrained list of candidate codewords, as the decoder tree is traversed, as opposed to only build the most likely codeword like the \gls{sc}-based algorithms. This List decoding algorithm was shown to significantly improve the error-correction performance compared to \gls{sc}-based algorithms. This improvement, however, comes at the cost of a much greater complexity.

\section{Proposed Blind-Detection Method}\label{sec:algo}

In this section, we describe a low-complexity algorithm that allows to discard most candidates before the higher-complexity subsequent decoder is executed. Our detection algorithm is based on the fast-SSC decoding algorithm where, alongside the decoding process, a detection metric is calculated. We propose a detection metric $\mathcal{D}$ where the update rules exploit the inherent structure of the various constituent codes. The bigger the value of $\mathcal{D}$, the more likely a noisy received message (block) \added{was encoded using} a polar code with the expected blocklength and code rate. The last step of the detection algorithm consists in selecting candidates with $\mathcal{D}$ greater than some predefined threshold.

We note that the detection metric proposed in the sequel has some similarities with the path metric used in list decoding of polar codes~\cite{Balatsoukas_TSP_2015,Sarkis_JSAC_2016}. However, the path metric used in list decoding is proportional to the likelihood of each estimated codeword given that a valid codeword was transmitted and given a noisy observation of that codeword. For blind detection of polar codes, on the other hand, the aim is to provide an estimate of the likelihood that a noisy channel observation was produced by a valid polar codeword. Thus, the proposed detection-metric update rules are modified with respect to the path metric update rules for list decoding in order to better fit the purpose of blind detection.

\subsection{Detection-Metric Update Rules}\label{sec:update-rules}
Following the same notation as in \cite{Sarkis_JSAC_2014}, $N_v$ designates the blocklength of a constituent code with its root at node $v$---in a decoder-tree representation---, and \glspl{llr} are denoted as $\alpha$. We use $\alpha_a^b$ to denote a vector of length $b-a+1$ and $\alpha_i$ is the $i^{\text{th}}$ element on the vector $\mvec{\alpha}$. We assume that positive and negative \glspl{llr} are mapped to 0 and 1, respectively. The detection metric $\mathcal{D}$ is initialized as $\mathcal{D}_0=0$.

\subsubsection{Rate-0 Code}
Entirely composed of frozen bits, rate-0 ``codes'' are not really codes, i.e., they are known a priori to be an all-zero vector. In a noiseless transmission, the \glspl{llr} to a rate-0 node shall be composed of all positive \glspl{llr}. Thus we propose the following update rule for $\mathcal{D}$:
\begin{equation}
  \mathcal{D}_{t}=\mathcal{D}_{t-1}+\frac{1}{N_v}\left( \sum_{i=0}^{N_v-1}{\alpha_i} \right).
  \label{eq:rate0}
\end{equation}

\textbf{Rationale}:
\begin{itemize}
  \item Even if the received vector $\alpha_0^{N_v-1}$ is noisy, a decodable frame should contain a majority of positive \glspl{llr} $\alpha_i$.
  \item If the input to this node is random, including if nothing was transmitted, the sum will average to zero.
\end{itemize}

\subsubsection{Rate-1 Code}
By definition, a rate-1 code does not contain any frozen bit, i.e., no redundancy is added to the information. This makes rate-1 codes useless for the purpose of detection and they are thus ignored in the calculation of the detection metric.

\subsubsection{Repetition Code}
A Repetition code is a code of rate $R_v=\nicefrac{1}{N_v}$ where an input is repeated $N_v$ times at the output. We propose the following update rule for $\mathcal{D}$:
\begin{equation}
  \mathcal{D}_{t}=\mathcal{D}_{t-1}+\frac{1}{N_v}\left| \sum_{i=0}^{N_v-1}{\alpha_i} \right|.
  \label{eq:rep}
\end{equation}

\textbf{Rationale}:
\begin{itemize}
  \item Even if the received vector $\alpha_0^{N_v-1}$ is noisy, a decodable frame should contain a majority of \glspl{llr} $\alpha_i$ that agree, i.e., share the same sign, and the amplitude of their sum should be greater than that of the wrong \glspl{llr}.
  \item If the input to this node is random, including if nothing was transmitted, the sum will average to zero (at least for sufficiently large $N_v)$.
\end{itemize}

It should be noted that the absolute value in the right-hand-side term renders this function non-negative. As a result, this update rule pushes $\mathcal{D}$ towards greater values as the amplitudes of $\alpha_i$ values increase with $\nicefrac{E_b}{N_0}$.

\subsubsection{SPC Code}\label{sec:algo:spc}
An \gls{spc} code is a constituent code of rate $R_v=\nicefrac{N_v-1}{N_v}$ where, after encoding, the least-significant bit location holds the parity of the $N_v-1$ information bits. We propose the following update rule for $\mathcal{D}$:

\begin{equation}
  \mathcal{D}_{t}=\mathcal{D}_{t-1}+(-1)^p\min\left( \left| \alpha_0^{N_v-1} \right| \right ),
  \label{eq:spc}
\end{equation}
where $p$ is the calculated parity based on hard decisions \cite[eq.\,(6)]{Sarkis_JSAC_2014} and $\min(\cdot)$ returns the smallest element of its input vector. Thus, the metric is increased when the parity is satisfied, and decreased otherwise.\\

\textbf{Rationale}:
\begin{itemize}
  \item Contrary to rate-0 or Repetition codes, an \gls{spc} code carries very little structural information. In fact, if an \gls{spc} node is fed random \glspl{llr}, the parity will be satisfied with probability $\nicefrac{1}{2}$.	 For this reason, it is the smallest of the absolute \gls{llr} values that is used to update the metric.
\end{itemize}

Using the defined detection metric and the corresponding update rules, the decision rule of our proposed detection algorithm for a given decision threshold $d$ can be written as
\begin{align}
	D\left(\mathcal{D}\right) & = \left\{ \begin{array}{cc} 
									\mathcal{H}_0, & \mathcal{D} < d, \\
									\mathcal{H}_1, & \mathcal{D} \geq d, \\
								\end{array}
								\right.
\end{align}
where $\mathcal{H}_0$ and $\mathcal{H}_1$ correspond to the null and alternate hypotheses, respectively.

\subsection{Complexity of the Detection Algorithm}
The proposed detection algorithm is based on fast-SSC decoding and, thus, its complexity is almost identical to that of a fast-SSC decoder with the only additional, but negligible, complexity of the update of the detection metric. However, it should be noted that this is the worst-case complexity as, in principle, it is not mandatory for the detector to fully decode each (potential) codeword since retained candidates will typically be fully decoded by the following module, e.g., a CRC-aided List decoder. Hence, the complexity of the detector could be significantly reduced by either only running the detection algorithm on a fraction of the received block or by introducing an early-stopping criterion that would, e.g., render its decision once a certain threshold has been met~\cite{Ghanaatian2016}.

\subsection{Limitations of the Detection Algorithm}
As already stated in Section~\ref{sec:algo:spc}, \gls{spc} codes contain very little structural information about a polar-encoded frame. Hence, we expect our proposed detection metric to become less and less reliable as the proportion of \gls{spc} codes in a polar code grows over the one of Repetition and rate-0 codes, which usually happens as the code rate is increased. To address this issue, at least three mitigation avenues could be explored: 1) constrain the maximum \gls{spc} node size, 2) only update the metric for a fraction of the total \gls{spc} nodes, 3) add a scaling factor to its metric update rule to attenuate its contribution.

\section{Simulation Results}\label{sec:results}
In this section, we provide simulation results that demonstrate the effectiveness of our detection algorithm. More specifically, we first evaluate the distribution of the detection metric under various transmission scenarios and then we focus on the detection and miss rates by showing the \gls{roc} of our detector.

\added{We assume that the low-complexity blind detector receives \glspl{llr} and that it passes the retained candidates to a complex decoder such as a CRC-aided List decoder with a list size $L=8$, the baseline decoding algorithm considered for the future 5G standard~\cite{3GPPRAN188}.}
All simulation results are for a \gls{bpsk} modulation used over an \gls{awgn} channel.

\subsection{Considered Transmission Scenarios}
For the simulation results, we consider the following transmission scenarios.
\subsubsection{No Transmission (NoTx)}
This is a scenario where no data was transmitted over the channel. Low values for the detection metric $\mathcal{D}$ are expected as the sums of both \eqref{eq:rate0} and \eqref{eq:spc} should average to 0 and, although non-negative, the contribution of \eqref{eq:rep} should be very small.

\subsubsection{Random Transmission (RndTx)}
This is a scenario where random data was transmitted over the channel. It simulates the case where the channel is being used but contains data that does not exhibit the structure inherent to the polar code to be detected.

\subsubsection{Regular Transmission (RegTx)}
Lastly, this scenario is for the case where frames encoded with the particular polar code of interest were transmitted over the channel. This scenario represents the case where the channel contains a polar coded block that should be detected in order to be passed on to a decoder.\vspace{5pt}

Following standard hypothesis testing nomenclature and notation, the union of the NoTx and RndTx scenarios forms the null hypothesis of our detector and is denoted by $\mathcal{H}_0$, while the RegTx scenario forms the alternate hypothesis and is denoted by $\mathcal{H}_1$.

\begin{figure}[t]
  \centering
  \includegraphics[width=0.65\columnwidth]{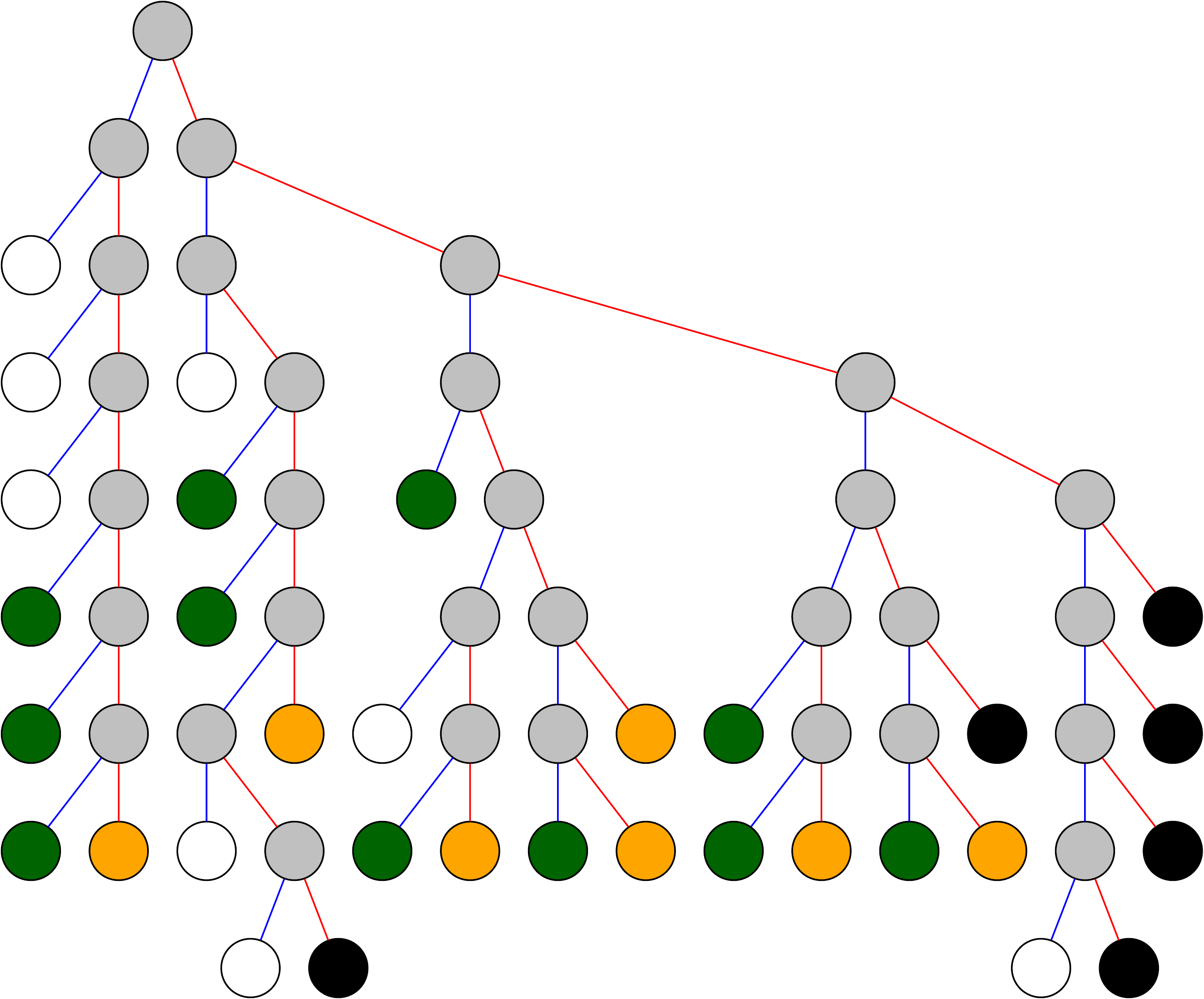}
  \caption{Decoder-tree representation of the $(512, 80)$ polar code used for the simulations.}
  \label{fig:decoder_tree}
\end{figure}

\subsection{Choice of Polar Code}
In order to provide meaningful and useful results for the next-generation downlink control channel which has not been finalized yet, we use some parameters from the existing LTE standard~\cite{LTE.TS36.211,LTE.TS36.213} as well as others derived from the current 3GPP RAN1 meeting discussions, e.g., as reported in \cite{3GPPRAN187,3GPPRAN188}. Hence, we assume that the length of the polar code protecting the control information messages will be short---maximum length $N_{\max}=512$---, and of low rate, e.g., a rate of $R=\nicefrac{1}{8}$ has often been discussed. We also assume that a \gls{crc} is always appended to messages, and that 16 bits is a typical length for the \gls{crc}.

The experimental results are given for a $(512, 80)$ systematic polar code optimized for an $\nicefrac{E_b}{N_0}$ of $2\,$dB, constructed using the method of Tal and Vardy~\cite{Tal2011a}. To give an idea of the constituent-code distribution, Fig.~\ref{fig:decoder_tree} illustrates this polar code in the form of a decoder tree, where rate-0, rate-1, Repetition, and \gls{spc} codes are shown as white, black, green, and orange nodes, respectively.

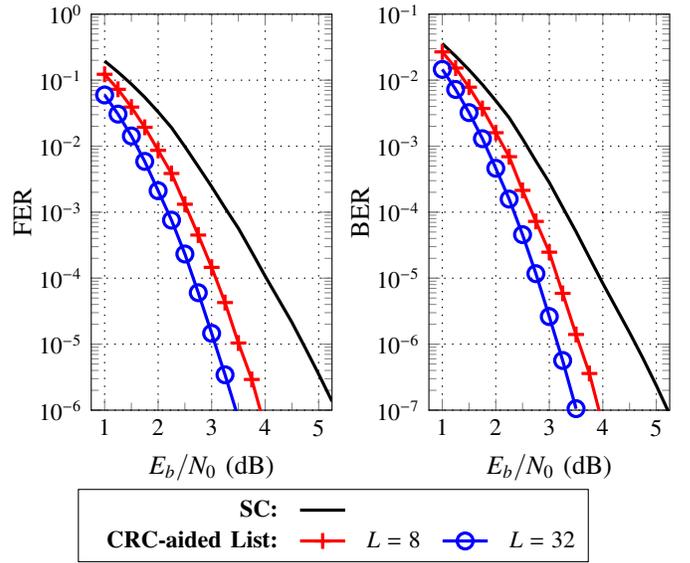
\begin{figure}[t]
  \centering
  \usetikzlibrary{plotmarks}

\begin{tikzpicture}

  \pgfplotsset{
    grid style = {
      dash pattern = on 0.05mm off 1mm,
      line cap = round,
      black,
      line width = 0.5pt
    },
    label style = {font=\fontsize{10pt}{7.2}\selectfont},
    tick label style = {font=\fontsize{9pt}{7.2}\selectfont}
  }

 \begin{semilogyaxis}[%
    xlabel=$E_b/N_0$ (dB),xtick={1,2,...,6.0},%
    xlabel style={yshift=0.2em},%
    minor x tick num={1},
    xmin=0.75,xmax=5.25,%
    ymin=1e-6,ymax=1e-0,%
    ylabel={FER}, ylabel style={yshift=-0.9em},%
    width=0.54\columnwidth, height=\ecperffigheight, grid=major,%
    legend style={
      anchor={center},
      cells={anchor=west},
      column sep=2mm,
      font=\fontsize{9pt}{7.2}\selectfont,
      mark size=3.0pt
    },
    legend columns=3,
    legend to name=perf-legend,
    mark size=3.0pt]
    
    \addlegendimage{empty legend}
    \addlegendentry[anchor=east]{\textbf{SC:}}

    \addplot[very thick,color=black] table[x=ebn0_db,y=FER] {data/512.64+16.awgn.s1.42.sc.float.csv};
    \addlegendentry{}
    \addlegendimage{empty legend}
    \addlegendentry{}

    \addlegendimage{empty legend}
    \addlegendentry[anchor=east]{\textbf{CRC-aided List:}}

    \addplot[very thick,color=red,mark=+] table[x=ebn0_db,y=FER] {data/512.64.awgn.s1.42.list.l8.c16.float.csv};
    \addlegendentry{$L=8$}

    \addplot[very thick,color=blue,mark=o] table[x=ebn0_db,y=FER] {data/512.64.awgn.s1.42.list.l32.c16.float.csv};
    \addlegendentry{$L=32$}

  \end{semilogyaxis}
\end{tikzpicture}
\begin{tikzpicture}

  \pgfplotsset{
    grid style = {
      dash pattern = on 0.05mm off 1mm,
      line cap = round,
      black,
      line width = 0.5pt
    },
    label style = {font=\fontsize{10pt}{7.2}\selectfont},
    tick label style = {font=\fontsize{9pt}{7.2}\selectfont}
  }

 \begin{semilogyaxis}[%
    xlabel=$E_b/N_0$ (dB),xtick={1,2,...,6.0},%
    xlabel style={yshift=0.2em},%
    minor x tick num={1},
    xmin=0.75,xmax=5.25,%
    ymin=1e-7,ymax=1e-1,%
    ylabel={BER}, ylabel style={yshift=-0.9em},%
    width=0.54\columnwidth, height=\ecperffigheight, grid=major,%
    mark size=3.0pt]
    
    \addplot[very thick,color=black] table[x=ebn0_db,y=BER] {data/512.64+16.awgn.s1.42.sc.float.csv};
    \addplot[very thick,color=red,mark=+] table[x=ebn0_db,y=BER] {data/512.64.awgn.s1.42.list.l8.c16.float.csv};
    \addplot[very thick,color=blue,mark=o] table[x=ebn0_db,y=BER] {data/512.64.awgn.s1.42.list.l32.c16.float.csv};

  \end{semilogyaxis}
\end{tikzpicture}
\\
\ref{perf-legend}
  \caption{Error-correction performance of a $(512, 80)$ systematic polar code under both SC-based and CRC-aided List decoding. For list decoding, the 16-bit CRC is stored among the 80 information bits.}
  \label{fig:ec-perf}
\end{figure}

Fig.~\ref{fig:ec-perf} shows the error-correction performance of the aforementioned $(512, 80)$ systematic polar code for reference. The performance in terms of \gls{fer} (left) and \gls{ber} (right) is illustrated for both \gls{sc}-based and CRC-aided List decoding algorithms. Curves for CRC-aided List decoding are for a maximum list size $L\in\{8,32\}$ and a 16-bit CRC. It is important to note that the 16-bit CRC is stored within the 80 information-bit locations making the effective rate of the system $R=\nicefrac{1}{8}$ in the cases where CRC-aided List decoding is used. 

From Fig.~\ref{fig:ec-perf}, it can be seen that the error-correction gap between \gls{sc} decoding and CRC-aided List decoding grows with $\nicefrac{E_b}{N_0}$. At a \gls{fer} of $10^{-4}$, this gap is approximately of 1\,dB between \gls{sc} decoding and 16-bit CRC-aided List decoding with $L=8$. Increasing the list size $L$ to 32 results in a coding gap of 1.35\,dB at the same \gls{fer}. Looking at a \gls{fer} of $10^{-5}$, the gaps increased further, reaching 1.2\,dB and 1.65\,dB for the same respective algorithms. Comparing both curves for CRC-aided List decoding, it can be seen that the gap between $L=8$ and $L=32$ remains virtually constant, at approximately 0.5\,dB, across all $\nicefrac{E_b}{N_0}$ values.

\subsection{Detection-Metric Distribution}
To be effective, a good detection metric has to increase significantly faster for a polar-encoded frame compared to a frame that only contains random data or noise. In order to evaluate the proposed detection method, we compare the \gls{cdf} of the decision metric under both the null hypothesis $\mathcal{H}_0$ and the alternate hypothesis $\mathcal{H}_1$. 

\begin{figure}[t]
  \centering
  \input{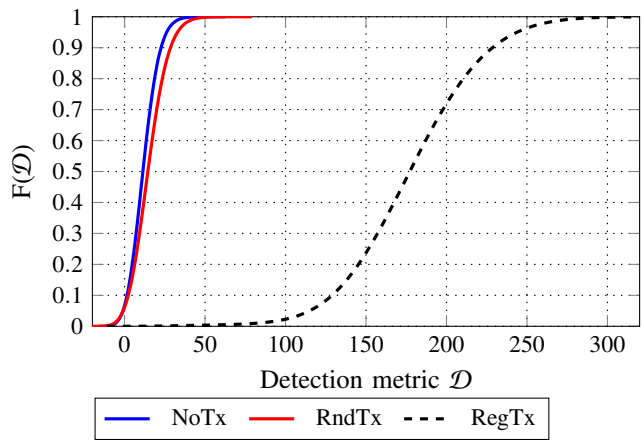}
  \caption{Comparison of the experimental \glspl{cdf} of $\mathcal{D}$, when no transmission occurs (NoTx) or random data was transmitted (RndTx) with the experimental \gls{cdf} when transmission of a valid polar-encoded frame occurs (RegTx). Results are for $\nicefrac{E_b}{N_0} = 3$\,dB.}
  \label{fig:metric-calc-NoTx-Vs-RndTx}
\end{figure}

The null hypothesis is a union of the NoTx and RndTx events, meaning that it is not possible to estimate the \gls{cdf} without knowing the prior distributions of these events. However, as can be seen in Fig.~\ref{fig:metric-calc-NoTx-Vs-RndTx}, when considering the \glspl{cdf} of the NoTx and RndTx events separately, we see that they are in fact very similar. Note that neither \gls{cdf} is centered around zero because, as pointed out in Section~\ref{sec:update-rules}, the update rule for the Repetition codes \eqref{eq:rep} is non-negative. Moreover, we observe in our simulations that the \gls{cdf} of the two events does not change significantly with the $\nicefrac{E_b}{N_0}$. For this reason, for the remaining comparison plots we use the worst-case \gls{cdf} among all our simulation results (i.e., the \gls{cdf} of RndTx for $\nicefrac{E_b}{N_0} = 3$\,dB) to avoid clutter.

The experimental \glspl{cdf} for $\mathcal{D}$ covering the scenarios of interest for various $\nicefrac{E_b}{N_0}$ values are shown in Fig.~\ref{fig:metric-calc-NoTx-Vs-RndTx}. We observe that, as can already be deduced from Fig.~\ref{fig:metric-calc-NoTx-Vs-RndTx}, the \glspl{cdf} for $\mathcal{D}$ under the null hypothesis $\mathcal{H}_0$ converge to $1$ much more quickly than under then alternate hypothesis $\mathcal{H}_1$. This shows that our proposed detection metric along with its update rules is a promising candidate for the purpose of blind detection of polar-encoded frames. Moreover, as the $\nicefrac{E_b}{N_0}$ is increased, the separation between the \glspl{cdf} becomes more apparent. We note that the $\nicefrac{E_b}{N_0}$ values were selected to approximately correspond to \glspl{fer} of $10^{-1}$, $10^{-2}$, $10^{-3}$, and $10^{-4}$ under 16-bit CRC-aided List decoding with $L=8$.

\begin{figure}[t]
  \vspace{-10pt}
  \centering
  \input{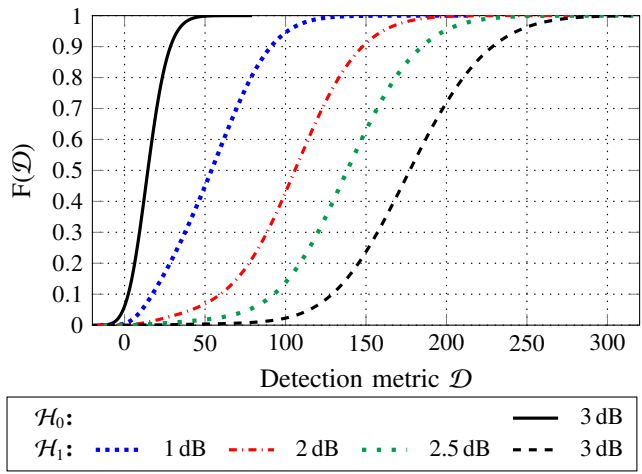}
  \caption{Comparison of the experimental cumulative distribution functions of $\mathcal{D}$ for $\mathcal{H}_0$ and $\mathcal{H}_1$. Results are for $\nicefrac{E_b}{N_0}=3$\,dB for $\mathcal{H}_0$ and $\nicefrac{E_b}{N_0} \in \{1,2,2.5,3\}$\,dB for $\mathcal{H}_1$.}
  \label{fig:metric-calc}
\end{figure}

\subsection{Detection Rate and Miss Rate}
In the previous section we saw that the distribution of the decision metric should enable reliable detection of polar-encoded frames. In this section, we quantify the performance of our proposed detector by plotting the miss probability as a function of the probability of false alarm. We note that this type of plot is very closely related to a \gls{roc} that is commonly used to characterize binary detectors.

The miss probability is usually defined as the probability of not detecting an event even though the event actually ocurred. In the case of our detector, this would correspond to the probability of not detecting a polar-encoded frame when a polar-encoded frame was, in fact, present. However, since our proposed detector will be used in conjunction with an actual polar decoder, it is more relevant to consider the probability of not detecting a polar-encoded frame that would have been decodable with the employed subsequent decoder. If we denote the event that a polar-encoded frame is present \emph{and} decodable by $\mathcal{F}_1$ and its complement by $\mathcal{F}_0$, then the miss and false alarm probabilities for a given detection threshold $d$ are given by 
\begin{align}	
	P_\text{miss}	& \triangleq \text{Pr}(\mathcal{D} < d\,|~\mathcal{F}_1),\\
	P_\text{fa}		& \triangleq \text{Pr}(\mathcal{D} \geq d\,|~\mathcal{F}_0),	
\end{align}
respectively.

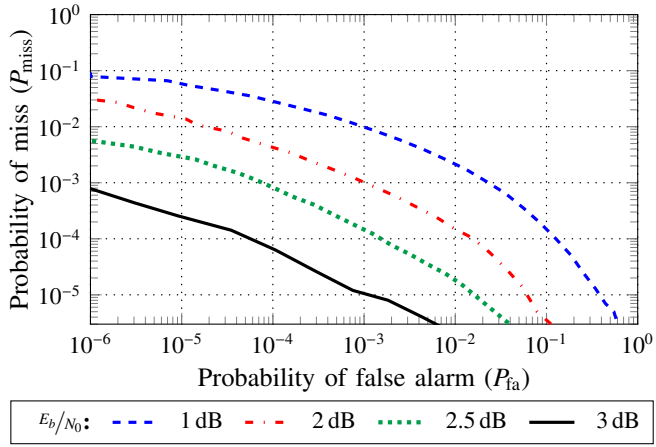
\begin{figure}
  \centering



\definecolor{greencb}{RGB}{0, 158, 75}

\begin{tikzpicture}

  \pgfplotsset{
    grid style = {
      dash pattern = on 0.05mm off 1mm,
      line cap = round,
      black,
      line width = 0.5pt
    },
    label style = {font=\fontsize{10pt}{7.2}\selectfont},
    tick label style = {font=\fontsize{9pt}{7.2}\selectfont}
  }

  \begin{axis}[%
    width=\columnwidth,
    height=\metricfigheight,
    xmin=1e-6, xmax=1,
    ymin=3e-6, ymax=1,
    yminorticks=true,
    grid = major,
		ylabel = {Probability of miss ($P_\text{miss}$)},
    ylabel style={yshift=-0.9em},
		xlabel = {Probability of false alarm ($P_\text{fa}$)},
    xlabel style={yshift=0.2em},
    xmode=log,
    ymode=log,
    legend style={
      anchor={center},
      cells={anchor=west},
      column sep=2mm,
      font=\fontsize{9pt}{7.2}\selectfont,
      mark size=3.0pt
    },
    legend columns=5,
    legend to name=roc-sc-legend,
    ]

    \addlegendimage{empty legend}
    \addlegendentry[anchor=east]{\textbf{$\nicefrac{E_b}{N_0}$:}}

    \addplot [color=blue, dashed, very thick]
    table[row sep=crcr]{%
      0.6528	2e-06\\
      0.547674	5e-06\\
      0.442278	7e-06\\
      0.343388	1.3e-05\\
      0.256766	2.4e-05\\
      0.185172	5.1e-05\\
      0.128936	9.70000000000001e-05\\
      0.086445	0.000188\\
      0.0562459999999999	0.00035\\
      0.0355969999999999	0.000641999999999999\\
      0.021904	0.001053\\
      0.013077	0.001736\\
      0.00756499999999999	0.002668\\
      0.004313	0.004043\\
      0.002411	0.005901\\
      0.001328	0.008405\\
      0.00073	0.011741\\
      0.000401	0.016016\\
      0.000199	0.021422\\
      0.000104	0.027839\\
      5.6e-05	0.03556\\
      2.6e-05	0.044394\\
      1.2e-05	0.054505\\
      7e-06	0.066022\\
      1e-06	0.078764\\
      1e-06	0.0927099999999999\\
    };
    \addlegendentry{1\,dB}

    \addplot [color=red, loosely dashdotted, very thick]
    table[row sep=crcr]{%
      0.112244	3e-06\\
      0.078902	5e-06\\
      0.054114	1.4e-05\\
      0.036154	2.9e-05\\
      0.023817	5.7e-05\\
      0.015453	0.000103\\
      0.00975999999999999	0.000152\\
      0.00599799999999999	0.000251\\
      0.003713	0.000373\\
      0.002228	0.000573\\
      0.000761999999999999	0.001235\\
      0.000450999999999999	0.001738\\
      0.000261	0.002435\\
      0.000156	0.003433\\
      8.59999999999999e-05	0.004602\\
      4.7e-05	0.006248\\
      3.2e-05	0.00825299999999999\\
      1.5e-05	0.010704\\
      1.1e-05	0.013826\\
      5e-06	0.017391\\
      3e-06	0.021827\\
      2e-06	0.027021\\
      1e-06	0.030818\\
    };
    \addlegendentry{2\,dB}

    \addplot [color=greencb, dotted, ultra thick]
    table[row sep=crcr]{%
      0.0568279999999999	2e-06\\
      0.03947	3e-06\\
      0.01193	1.5e-05\\
      0.007771	2.4e-05\\
      0.00498999999999999	3.5e-05\\
      0.003176	5.3e-05\\
      0.001998	7.70000000000001e-05\\
      0.00123	0.000124\\
      0.000766	0.00018\\
      0.000464	0.000269\\
      0.000284	0.000404\\
      0.000163	0.000567999999999999\\
      0.000104	0.000802\\
      6.40000000000001e-05	0.001135\\
      4.2e-05	0.001484\\
      2.3e-05	0.001985\\
      1.4e-05	0.002624\\
      6e-06	0.003376\\
      3e-06	0.004408\\
      1e-06	0.005649\\
    };
    \addlegendentry{2.5\,dB}

    \addplot [color=black, very thick]
    table[row sep=crcr]{%
      0.043694	1e-06\\
      0.00994099999999999	2e-06\\
      0.004362	4e-06\\
      0.001832	8e-06\\
      0.000752999999999999	1.2e-05\\
      0.000299	2.6e-05\\
      0.000107	6.30000000000001e-05\\
      3.5e-05	0.000142\\
      9.00000000000001e-06	0.000258\\
      3e-06	0.000441\\
      1e-06	0.000783999999999999\\
    };
    \addlegendentry{3\,dB}
    
  \end{axis}
\end{tikzpicture}%
\\
\ref{roc-sc-legend}

  \caption{Receiver operating characteristic for the proposed detection metric $\mathcal{D}$ under SC decoding. Results are for $\nicefrac{E_b}{N_0} \in \{1,2,2.5,3\}$\,dB.}
  \label{fig:rocsc}
\end{figure}

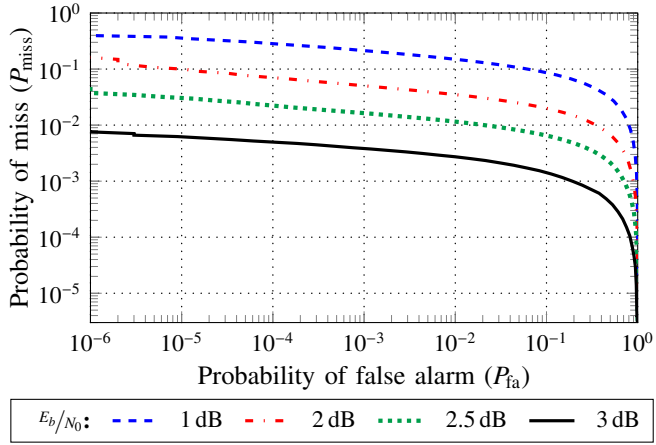
\begin{figure}
  \vspace{-5pt}
  \centering



\definecolor{greencb}{RGB}{0, 158, 75}

\begin{tikzpicture}

  \pgfplotsset{
    grid style = {
      dash pattern = on 0.05mm off 1mm,
      line cap = round,
      black,
      line width = 0.5pt
    },
    label style = {font=\fontsize{10pt}{7.2}\selectfont},
    tick label style = {font=\fontsize{9pt}{7.2}\selectfont}
  }

  \begin{axis}[%
    width=\columnwidth,
    height=\metricfigheight,
    xmin=1e-6, xmax=1,
    ymin=3e-6, ymax=1,
    yminorticks=true,
    grid = major,
		ylabel = {Probability of miss ($P_\text{miss}$)},
    ylabel style={yshift=-0.9em},
		xlabel = {Probability of false alarm ($P_\text{fa}$)},
    xlabel style={yshift=0.2em},
    xmode=log,
    ymode=log,
    legend style={
      anchor={center},
      cells={anchor=west},
      column sep=2mm,
      font=\fontsize{9pt}{7.2}\selectfont,
      mark size=3.0pt
    },
    legend columns=5,
    legend to name=roc-scl-legend,
    ]

    \addlegendimage{empty legend}
    \addlegendentry[anchor=east]{\textbf{$\nicefrac{E_b}{N_0}$:}}

    \addplot [color=blue, dashed, very thick]
    table[row sep=crcr]{%
      0.999982000000001	3e-06\\
      0.998406000000001	8.2e-05\\
      0.995226	0.000226\\
      0.987496000000001	0.000581\\
      0.971367000000001	0.001347\\
      0.942444999999999	0.002701\\
      0.897036	0.004936\\
      0.832836	0.008316\\
      0.749958	0.013146\\
      0.6528	0.019231\\
      0.547674	0.026795\\
      0.442278	0.035514\\
      0.343388	0.045289\\
      0.256766	0.055801\\
      0.185172	0.06695\\
      0.128936	0.078586\\
      0.086445	0.0904599999999999\\
      0.0562459999999999	0.10272\\
      0.0355969999999999	0.115368\\
      0.021904	0.128525\\
      0.013077	0.142438\\
      0.00756499999999999	0.157003\\
      0.004313	0.172332\\
      0.002411	0.188357\\
      0.00073	0.223163\\
      0.000401	0.242137\\
      0.000104	0.28258\\
      5.6e-05	0.304214\\
      1.2e-05	0.349079\\
      7e-06	0.372503\\
      1e-06	0.396468\\
      1e-06	0.420869\\
    };
    \addlegendentry{1\,dB}

    \addplot [color=red, loosely dashdotted, very thick]
    table[row sep=crcr]{%
      0.99998	1e-06\\
      0.997024	2.7e-05\\
      0.983958	0.000145\\
      0.968184	0.000288\\
      0.942586	0.000543999999999999\\
      0.904457	0.000903999999999999\\
      0.852106	0.001459\\
      0.786276999999999	0.002263\\
      0.708103	0.003313\\
      0.620867999999999	0.004669\\
      0.529729	0.006161\\
      0.438408	0.00792499999999999\\
      0.352909	0.009935\\
      0.275921	0.012046\\
      0.209823	0.014307\\
      0.155182	0.016689\\
      0.112244	0.019151\\
      0.078902	0.021598\\
      0.054114	0.024192\\
      0.036154	0.026851\\
      0.023817	0.029564\\
      0.015453	0.032341\\
      0.00975999999999999	0.035163\\
      0.003713	0.041416\\
      0.000450999999999999	0.0566509999999999\\
      0.000261	0.061139\\
      0.000156	0.066237\\
      4.7e-05	0.077586\\
      3.2e-05	0.0841639999999999\\
      1.5e-05	0.091204\\
      1.1e-05	0.0989739999999999\\
      5e-06	0.107288\\
      3e-06	0.116435\\
      2e-06	0.126384\\
      2e-06	0.148746\\
      1e-06	0.161214\\
    };
    \addlegendentry{2\,dB}

    \addplot [color=greencb, dotted, ultra thick]
    table[row sep=crcr]{%
      0.999435999999999	1e-06\\
      0.996021000000001	8e-06\\
      0.991223999999999	2.4e-05\\
      0.982135	5.3e-05\\
      0.966709999999999	0.0001\\
      0.942716	0.000166\\
      0.907877	0.000286\\
      0.861072000000001	0.000419\\
      0.802175	0.000626999999999999\\
      0.732519	0.000896999999999999\\
      0.653797000000001	0.001271\\
      0.569944	0.001725\\
      0.484598	0.002251\\
      0.402341	0.002803\\
      0.326091	0.003429\\
      0.257715	0.004079\\
      0.199231	0.004782\\
      0.150332	0.00549699999999999\\
      0.110888	0.006265\\
      0.079941	0.007046\\
      0.0568279999999999	0.007804\\
      0.03947	0.008616\\
      0.026904	0.009445\\
      0.018064	0.010299\\
      0.01193	0.011168\\
      0.007771	0.012064\\
      0.00498999999999999	0.012981\\
      0.00123	0.015924\\
      0.000163	0.020668\\
      0.000104	0.02219\\
      6.40000000000001e-05	0.023772\\
      4.2e-05	0.025522\\
      2.3e-05	0.027444\\
      1.4e-05	0.029545\\
      6e-06	0.031957\\
      3e-06	0.034603\\
      1e-06	0.037499\\
      1e-06	0.052952\\
    };
    \addlegendentry{2.5\,dB}

    \addplot [color=black, very thick]
    table[row sep=crcr]{%
      0.997658000000001	1e-06\\
      0.989501	4e-06\\
      0.980273	4e-06\\
      0.965206	1.4e-05\\
      0.942812	3e-05\\
      0.911202	4.6e-05\\
      0.869500000000001	6.90000000000001e-05\\
      0.817175	0.000107\\
      0.754436	0.00015\\
      0.684391	0.000212\\
      0.608433	0.000285\\
      0.529863	0.00038\\
      0.451689	0.000484\\
      0.377538	0.000608\\
      0.246823	0.000847999999999999\\
      0.19359	0.000993999999999999\\
      0.148915	0.001164\\
      0.112628	0.001344\\
      0.083556	0.001536\\
      0.060892	0.001718\\
      0.043694	0.001923\\
      0.030882	0.002118\\
      0.021441	0.002334\\
      0.014734	0.002524\\
      0.00994099999999999	0.002725\\
      0.00665999999999999	0.002906\\
      0.004362	0.003122\\
      0.00285	0.003333\\
      0.001194	0.003746\\
      0.000752999999999999	0.00395\\
      0.000299	0.004457\\
      0.000185	0.004708\\
      6.1e-05	0.00520899999999999\\
      3.5e-05	0.00552299999999999\\
      9.00000000000001e-06	0.006244\\
      3e-06	0.006615\\
      3e-06	0.00705899999999999\\
      1e-06	0.00757499999999999\\
      1e-06	0.00812\\
    };
    \addlegendentry{3\,dB}
    
  \end{axis}
\end{tikzpicture}%
\\
\ref{roc-scl-legend}

  \caption{Receiver operating characteristic for the proposed detection metric $\mathcal{D}$ under 16-bit CRC-aided List decoding with $L=8$. Results are for $\nicefrac{E_b}{N_0} \in \{1,2,2.5,3\}$\,dB.}
  \label{fig:roclist}
\end{figure}

In Fig.~\ref{fig:rocsc}, we present $P_\text{miss}$ as a function of $P_\text{fa}$ for our proposed detector for various $\nicefrac{E_b}{N_0}$ values when only an \gls{sc} decoder is used after the detector. As both probabilities are generally small, contrary to a traditional \gls{roc}, we use a logarithmic scale on both axes. We observe that, similarly to the previous section, as the $\nicefrac{E_b}{N_0}$ is increased the detector clearly becomes more effective. In particular, we see that for an $\nicefrac{E_b}{N_0}$ of $3$\,dB our detector can achieve a miss probability of $10^{-5}$ with a probability of false alarm as low as $10^{-3}$.

In Fig.~\ref{fig:roclist}, we present $P_\text{miss}$ as a function of $P_\text{fa}$ for our proposed detector for various $\nicefrac{E_b}{N_0}$ values when a 16-bit CRC-aide List decoder with $L=8$ is used after the detector. In this case, we observe that the performance of the detector is worse compared to the case where a \gls{sc} decoder follows the detector. This happens because a significantly higher fraction of undetected frames are in fact decodable, since the List decoding algorithm is more powerful than the \gls{sc} decoding algorithm. Thus, when a more powerful decoding algorithm is used, a more powerful detection algorithm should also be used in order to preserve the decoding capability of the high-performance decoder with the detection.

\section{Conclusion}\label{sec:conclusion}
In this paper, we proposed an algorithm for the blind detection of polar-encoded frames. The results show that our detection metric allows to distinguish polar-encoded frames from noisy received messages with great accuracy. The key ingredients are the update rules that exploit the inherent structure of constituent codes that compose a polar code. Our results indicate that our proposed detection metric, update rules, and algorithm are promising candidates for the implementation of a blind detector that would quickly reduce a list of potentially polar-encoded frame candidates to a manageable number.

\bibliographystyle{IEEEtran}
\bibliography{IEEEabrv,refs}

\end{document}